\documentclass[aps,prd,twocolumn,groupedaddress,nofootinbib,showpacs,preprintnumbers]{revtex4} 
\usepackage{graphics}
\usepackage{amsmath,amssymb,bm,graphicx,url,epsfig}
\usepackage{amsbsy}

\newcommand{\rf}[1]{(\ref{#1})}
\newcommand{\beq}{\begin{equation}}
\newcommand{\eeq}{\end{equation}}
\newcommand{\be}{\begin{equation}}
\newcommand{\ee}{\end{equation}}
\newcommand{\bea}{\begin{eqnarray}}
\newcommand{\eea}{\end{eqnarray}}
\newcommand{\eq}[1]{Eq.~(\ref{#1})}
\newcommand{\non}{\nonumber \\*}
\newcommand{\ie}{{i.e.}\ }
\newcommand{\brho}{\bar\rho}
\newcommand{\blambda}{\bar\lambda}
\newcommand{\LA}{\left\langle}
\newcommand{\RA}{\right\rangle}
\newcommand{\e}{\mbox{e}}
\renewcommand{\d}{\mbox{d}}

\newcommand{\om}{\omega}

\newcommand{\tr}{\mathrm{tr}\,}

\def\gtrsim{\mathrel{\mathpalette\fun >}}
\def\fun#1#2{\lower3.6pt\vbox{\baselineskip0pt\lineskip.9pt
\ialign{$\mathsurround=0pt#1\hfil##\hfil$\crcr#2\crcr\sim\crcr}}}

\def\ga{\gtrsim} 

\begin{document}


\title{Stability of the nonperturbative bosonic string vacuum}

\author
{J. Ambj\o rn$\,^{a,b}$ and Y. Makeenko$\,^{a,c}$}

\affiliation{\vspace*{2mm}
${}^a$\/The Niels Bohr Institute, Copenhagen University,
Blegdamsvej 17, DK-2100 Copenhagen, Denmark\\
${}^b$\/IMAPP, Radboud University, Heyendaalseweg 135,
6525 AJ, Nijmegen, The Netherlands\\
${}^c$\/Institute of Theoretical and Experimental Physics,
B. Cheremushkinskaya 25, 117218 Moscow, Russia\\
	\vspace*{1mm}
{email: ambjorn@nbi.dk \ makeenko@nbi.dk}
}


\begin{abstract}
Quantization of the bosonic string around the classical, perturbative vacuum
is not consistent  for  spacetime  dimensions $2<d<26$.
Recently we have showed that at large $d$ there is another so-called mean-field 
vacuum. Here we extend
 this mean-field calculation to finite $d$ and show that 
 the corresponding mean-field vacuum is 
stable under quadratic fluctuations for $2 < d < 26$. 
We point out the analogy with the two-dimensional  $O(N)$-symmetric sigma-model,
where the $1/N$-vacuum is very close to the real vacuum state even for
finite $N$, in contrast to the perturbative vacuum.
\end{abstract}

\pacs{11.25.Pm, 11.15.Pg,} 


\maketitle

\section{Introduction}

The action of the Nambu-Goto string is the area of the string world sheet. It is 
highly nonlinear in the embedding-space coordinates. Making use of diffeomorphism 
invariance and fixing a gauge makes the action quadratic but nonlinearities are now hidden
in the dependence of the cutoff on the metric induced at the world sheet.

If one uses the Polyakov formulation~\cite{Pol81} of string theory,  the embedding-space
 coordinates and the intrinsic world sheet metric
are independent. The  action is quadratic in the embedding-space coordinates and 
in the path integral one can in principle perform the integration over these 
coordinates. The dependence of  this part of the path integral on the world sheet 
metric is determined by the conformal anomaly. In the conformal 
gauge this leads to the the celebrated Liouville action whose solution~\cite{KPZ} about the 
classical (perturbative) vacuum is consistent only for $d\leq2$.
For $2 < d <26$ the solution is not real which 
may indicate an instability of the 
vacuum.

In the work~\cite{AM15} we constructed another  vacuum state of the Nambu-Goto 
string by introducing an independent intrinsic metric $\rho_{ab}$ and the 
corresponding Lagrange multiplier $\lambda^{ab}$ and then integrating   
over the $d$ target-space coordinates $X^\mu$. The corresponding effective action is a functional of $\rho_{ab}$ and $\lambda^{ab}$ which do not fluctuate in the mean-field approximation that becomes exact at large $d$.
A vacuum state can be found by minimizing the effective action and it is a 
genuine quantum state because we have taken
into account the quantum fluctuations of the target-space coordinates $X^\mu$. 

This approach is quite similar to the well-known introduction
of a Lagrange multiplier field in the two-dimensional $O(N)$ sigma-model.
In this model  one integrates over  fields
$\vec n$ obeying  the restriction $\vec n^2=1$. One gets rid of the constraint 
by introducing the Lagrange multiplier field $u$.
After integration over $\vec n$ one obtains an effective action as a functional of $u$.
The minimum of this effective action determines  
the exact vacuum state for  infinite  $N$. For  finite $N$ the quantum fluctuations of $u$ have 
to be included, but they are small even at $N=3$. The reason is roughly speaking 
that there is only one $u$, while the effective action is proportional to $ N$, \ie
parametrically large, which is what is needed for a saddle point. 
That is  not the case for the $N$ fields $\vec n$ which fluctuate strongly.
The perturbative vacuum $\vec n_{\rm cl}=(1,0,\ldots,0)$  possesses an $O(N-1)$ symmetry and is far away from the genuine nonperturbative $O(N)$-symmetric vacuum, 
while the mean-field vacuum obtained via the Lagrange multiplier approach 
possesses the right symmetry and is close to the exact vacuum  even at finite $N$. 
Fluctuations of $u$ about the mean-field value are
systematically tractable within the $1/N$-expansion.

In the present Paper we construct  a nonperturbative  mean-field 
vacuum state  for the Nambu-Goto string at finite $d$, show that it is energetically preferable
to the parturbative classical vacuum and discuss two possible scaling limits.
We calculate the effective action which governs 
fluctuations of  $\rho_{ab}$ and $\lambda^{ab}$ 
about their mean-field values,  
repeating pretty much the original computation
of the conformal anomaly in Ref.~\cite{Pol81} for the Polyakov string. 
Fixing the conformal gauge we evaluate the determinants coming from path-integration over $X^\mu$ and ghosts,
in order to compute the effective action to
quadratic order in $\delta\rho_{ab}$ and $\delta \lambda^{ab}$ at finite $d$.
We show that it is positive definite for $2<d<26$, but becomes unstable 
in the stringy scaling limit for $d>26$.

\section{The mean-field vacuum\label{s:mf}}

Let us consider a closed bosonic string in a target space with one  compactified dimension
of circumference $\beta$,  whose world sheet wraps once around
this compactified dimension. There is no tachyon with this setup
if $\beta$ is sufficient large for the classical energy squared to be larger than (minus) the tachyon mass
squared.
The Nambu-Goto action is given by the area of the embedded 
surface which we rewrite using a Lagrange multiplier $\lambda^{ab}$
and an independent intrinsic metric $\rho_{ab}$ as
\bea
\lefteqn{
K_0 \!\int \d^2\omega\,\sqrt{\det \partial_a X \cdot \partial_bX}=
K_0 \!\int \d^2\omega\,\sqrt{\det  \rho}   } \non &&
+\frac{K_0}2 \!\int \d^2\omega\, \lambda^{ab} \left( \partial_a X \cdot \partial_bX -\rho_{ab}
\right).~~~~~~~~~~
\label{aux}
\eea
We perform quantization by
the path integral which goes over real $X^\mu(\omega)$ and $\rho_{ab}(\omega)$ and
imaginary $\lambda^{ab}(\omega)$.
We choose the world sheet coordinates  $\omega_1$ and $\omega_2$ inside 
a $\omega_L\times \omega_\beta$ rectangle in the parameter space, when
the classical solution $X^\mu_{\rm cl}$ minimizing the action \rf{aux} is $\omega$-independent.

We integrate out quantum fluctuations of the fields $X^\mu$ by splitting
$X^\mu=X^\mu_{\rm cl} +X^\mu_{\rm q}$ and then  performing  the Gaussian path
integral over $X^\mu_{\rm q}$. 
We thus obtain the effective action
governing the fields $\lambda^{ab}$ and $\rho_{ab}$,
\bea
S_{\rm eff}&=& K_0 \int \d^2\omega\,\sqrt{\det \rho} 
+\frac{K_0}2\! \int \d^2\omega\, \lambda^{ab} \left( \partial_a X_{\rm cl } 
\cdot \partial_bX_{\rm cl }  \right.  \non && \left.
 -\rho_{ab} \right) 
+ \frac{d}{2}  \tr \log (-{\cal O}),  \non 
{\cal O}&  := &\frac1 {\sqrt{\det \rho} }  \partial _a \lambda^{ab} \partial_b.
\label{aux1}
\eea
The operator ${\cal O}$
 reproduces the usual two-dimensional  Laplacian for  $\lambda^{ab}= \rho^{ab}\sqrt{\det\rho}$.

The action \rf{aux1} is the effective action for path-integration over $\rho_{ab}$ and
$\lambda^{ab}$. Making use of diffeomorphism invariance one can choose the
conformal gauge, diagonalizing $\rho_{ab}=\rho \delta_{ab}$. This  
produces the ghost determinant~\cite{Pol81}
\be
{\cal D}\rho_{ab} = {\cal D}\rho \det(-{\cal O}_{\rm gh}).
\ee
Here the operator
\bea
{\cal O}_{\rm gh}&:=&
\left(\Delta-\frac12 R\right)\delta^a_b=
\left[ \frac 1{\rho} \partial^2 -\frac 1{2\rho} (\partial^2 \log \rho) \right]\delta^a_b ~~~
\label{gh}
\eea
acts on 2D vector functions obeying the mixed boundary conditions: Dirichlet for one component
and Robin for the other. 
This produces the term 
\be
S_{\rm gh}=-\tr \log (-{\cal O}_{\rm gh})
\ee
to be added to the right-hand side of \eq{aux1}.

It is easy to compute the determinants for constant fields $\rho_{ab}=\brho\delta_{ab}$
and $\lambda^{ab}=\blambda \delta^{ab}$. We may consider these as an ansatz for
the values of $\rho_{ab}$ and $\lambda^{ab}$ minimizing the effective
action. We shall then demonstrate that quadratic fluctuations around this minimum
are stable, so it is indeed a solution minimizing the effective action.

Computing the determinants for constant  $\rho_{ab}=\brho\delta_{ab}$
and $\lambda^{ab}=\blambda \delta^{ab}$, we obtain for $L\gg\beta$
the well-known result
\bea
S_{\rm eff}+S_{\rm gh} 
&=&\frac {K_0}2 \blambda \left( \frac {L^2}{\omega_L^2}+ \frac {\beta^2}{\omega_\beta^2}
\right)\omega_L\omega_\beta -\frac{\pi(d-2)}{6} \frac{\omega_L}{\omega_\beta} \non
&& + \left(K_0 -K_0\blambda-\frac{d\Lambda^2}{2\blambda}+
 \Lambda^2\right)\brho\,\omega_L\omega_\beta,
\label{SpS}
\eea
where $\Lambda^2$ cuts off eigenvalues of the operators involved
 (which
are parametrization-independent), the cutoff of integration over the proper time
being precisely $(4\pi \Lambda^2)^{-1}$.

The minimum of \rf{SpS}
with respect to $\brho$, $\blambda$ and $\omega_\beta$ is reached at
\begin{subequations}
\bea
\blambda&=&C:= \frac{1}2 +\frac{\Lambda^2}{2K_0} +
\sqrt{\left(\frac{1}2 +\frac{\Lambda^2}{2K_0}\right)^2 -\frac{d\Lambda^2}{2K_0}},~~
\label{newC} \\
\brho&=& \frac{L}{\omega_L\omega_\beta}
\frac{\left(\beta^2-\frac{\pi(d-2)}{6K_0 C}\right)}
{\sqrt{\beta^2-\frac{\pi(d-2)}{3K_0 C}}}
\frac C{\left(2C-1-\frac {\Lambda^2}{K_0}\right) },
\label{newrho} \\
\omega_\beta&=& \frac{\omega_L}{L}\sqrt{\beta^2-\frac{\pi(d-2)}{3K_0 C}}.
\eea
\label{mmff}
\end{subequations}
The value of \rf{SpS} at the minimum determines the energy of the ground state
\be
E_0 = K_0 C \sqrt{\beta^2-\frac{\pi(d-2)}{3K_0 C}}.
\label{Sfin}
\ee
It is explicitly seen from this formula that the energy is not tachyonic if $\beta$ is large enough
for the difference under the square root to be positive.

The solution \rf{mmff}, \rf{Sfin} reproduces the one of Ref.~\cite{AM15} as $d\to \infty$
(when $K_0\sim d$) and generalizes it to finite $d$.
Equations \rf{mmff} describe a nonperturbative vacuum 
in the mean-field approximation,  where
we disregard fluctuations
of  $\lambda^{ab}$ and $\rho_{ab}$ about the  saddle-point values
$\blambda^{ab}$ and $\brho_{ab} $.
Note that $C$ as given in \rf{newC} takes values between 
1 and 
\be
C_*=\frac{1}{2} \left(d-\sqrt{d^2-2 d}\right)
\label{C*}
\ee
(monotonically changing from 1/2 to 1 with $d$ decreasing from $\infty$ to 2)
as $K_0$ decreases from infinity to 
\be
K_*=\left(d-1+\sqrt{d^2-2d}\right) \Lambda^2.
\label{K*}
\ee
$C$ would start from its classical value 1, 
if one were performed a perturbative expansion in $1/K_0$,
\ie about the classical (perturbative) vacuum.
This is also true for \rf{newC} and
\rf{newrho}  which would start out with their classical values.
However, as described in Sect.~\ref{s:scaling}
the continuum nonperturbative vacuum is approached as 
$K_0\to K_*$ and correspondingly $ C\to C_*$.

\section{Instability of the classical vacuum}

It is clear that the ground-state energy \rf{Sfin} is always smaller for $d>2$ than
its classical value 
$K_0 \beta$  because $C<1$. For this reason the mean-field vacuum~\rf{mmff}
is energetically preferable to the perturbative, classical vacuum which is thus unstable.

To understand this instability,
it is instructive to compute an ``effective potential'', like in the studies of 
symmetry breaking in quantum field theory. For this purpose we add to the action \rf{aux}
the source term
\be
S_{\rm src}=\frac {K_0}2 \int \d^2 \omega \, j^{ab} \rho_{ab}
\ee
and define the partition function $Z[j]$ in the presence of the source by path integration over
the fields. 
Repeating easily the above mean-field computation for constant $j^{ab}=j \delta^{ab}$, we find
\be
\brho(j) 
\equiv -\frac 1{K_0 L\beta}\frac {\partial \log Z[j]}{\partial j}
=\frac12+
\frac { 1+j+\frac{\Lambda^2}{K_0} }{\sqrt{\left(1 +j+\frac{\Lambda^2}{K_0}\right)^2 -\frac{2d\Lambda^2}{K_0}}}
\label{rhoj}
\ee
for $\omega_L=L$ and $\omega_\beta=\beta \gg 1\sqrt{K_0}$,  reproducing then \rf{newrho}  for $j=0$. 

The effective potential $\Gamma(\brho)$ is defined in the standard way 
by the Lagrange transformation
\be
\Gamma[\brho]\equiv -\frac 1{K_0 L\beta}
\left(\log Z[j] +\frac{K_0}2\int \d^2 \omega\, j^{ab} \brho_{ab}(j) \right).
\ee
Solving \eq{rhoj} for $j$ we obtain
\be
j(\brho)=-1-\frac{\Lambda^2}{K_0}+\sqrt{\frac{d\Lambda^2}{2K_0}}
\frac{(2\brho-1)}{ \sqrt{\brho(\brho-1)}} ,
\label{jjrr}
\ee
which results in
\be
\Gamma(\brho)=
\left( 1+\frac{\Lambda^2}{K_0} \right) \brho -\sqrt{\frac{2d \Lambda^2}{K_0}  \brho(\brho-1)}
\label{Ga}
\ee
in the mean-field approximation. Note that
\be
-\frac{\partial \Gamma(\brho)}{\partial \brho}=j(\brho)
\label{10}
\ee
with $j(\brho)$ given by \eq{jjrr} as it should.

Near the classical  vacuum when $0<\brho-1\ll 1$ the potential
\rf{Ga} decreases with increasing
$\brho$ because of the second term with the negative sign, which demonstrates an 
instability of the classical vacuum. If 
$K_0>K_* $ given by \eq{K*},  
the potential \rf{Ga} linearly increases with $\brho$ for large $\brho$ and thus has a (stable) minimum at 
\be
\brho(0) =\frac 12 + \frac{1+\frac{\Lambda^2}{K_0}}
{2\sqrt{\left(1+\frac{\Lambda^2}{K_0}\right)^2-\frac {2d\Lambda^2}{K_0}}}
\label{bbrr}
\ee
which is the same as \rf{newrho} for $\beta\gg1/\sqrt{K_0}$.
Near the minimum we have
\bea
\Gamma(\brho)&=&C+  \frac{K_0}{2d \Lambda^2} 
\left[ \left(1+\frac{\Lambda^2}{K_0}\right)^2-\frac {2d\Lambda^2}{K_0} \right]^{3/2}\!
\left[ \brho-\brho(0) \right]^2  \non &&+
{\cal O}\left(\left[ \brho-\brho(0) \right]^3\right).
\eea
The coefficient in front of the quadratic term is positive for $K_0>K_*$ which explicitly
demonstrates stability of the minimum.

We thus conclude that the effective potential $\Gamma(\brho)$ is lower for
the (stable) mean-field minimum \rf{bbrr} than for the 
perturbative, classical vacuum $\brho=1$. The latter is therefore unstable. It looks
like a dynamical symmetry breaking in quantum field theory that generates
a nontrivial world sheet metric \rf{bbrr}. This also determines 
the 
averaged induced metric because
\be
\LA \partial _a X \cdot \partial _b X \RA =\brho_{ab}
\ee
in the mean-field approximation.

\section{Scaling limit and renormalization\label{s:scaling}}

The renormalization of 
 the formulas \rf{newrho}, \rf{Sfin}  can be performed
quite similarly to the one discussed in 
Ref.~\cite{AM15} where we had $K_*=2d\Lambda^2 $ and $C_*=1/2$ at large $d$.
In \cite{AM15} we discussed two possibilities for renormalization, one led to 
what we called  ``Gulliver's world", and it  is  the renormalization one has 
been using  when  one regularized the string theory on a hyper-cubic  lattice 
\cite{GT77,lattice} or via dynamical  triangulations \cite{DT} in $d$ dimensions. 
The other possibility led to 
what we denoted the ``Lilliputian world'', and it is the renormalization where we 
reproduce some of the standard continuum string theory results. 

In both cases we define a renormalized string tension $K_R$ by 
\be
K_R= K_0 
{\sqrt{\left( 1+\frac{\Lambda^2}{K_0}  \right)^2-\frac{2d\Lambda^2}{K_0}}}
=K_0 \left(2C-1-\frac {\Lambda^2}{K_0} \right)
\label{KR}
\ee  
and insist that it stays finite in the limit $\Lambda \to \infty$. This requirement 
corresponds to the following scaling behavior of $K_0$ for $\Lambda \to \infty$:
\be
K_0\to K_*+\frac{K_R^2}{2\Lambda^2\sqrt{d^2-2d}}.
\label{sca}
\ee
With this scaling we have for $\Lambda \to \infty$ that 
\be\label{xj2}
\left(K_0-\frac{d\Lambda^2}{2C^2}\right)\to \frac{K_R}{C_*} ,
\qquad 2C -1 - \frac{\Lambda^2}{K_0} \to 
\frac{K_R}{K_*},
\ee
where  $C_*$ and $K_*$ given by Eqs.~\rf{C*} and  \rf{K*} are positive functions of $d$ for $ 2<d<\infty$.

As described in \cite{AM15} the difference between the ``Gulliver" and the ``Lilliputian" 
renormalizations was that in the lattice approach we did not have the freedom
to perform further renormalization, while in the ``Lilliputian" case we could perform
an additional ``background field'' renormalization of the external lengths $L$ and $\beta$: 
\be
L_R= L\sqrt{\frac C{2C-1-\frac {\Lambda^2}{K_0} } } , \quad 
\beta_R=\beta \sqrt{\frac C{2C-1-\frac {\Lambda^2}{K_0} } } .
\label{LR}
\ee
By insisting that $L_R$ and $\beta_R$ remain finite when $\Lambda\to\infty$, rather 
than $L$ and $\beta$ do as in lattice string theory, it follows from \rf{xj2} that 
$L$ and $\beta$ go to zero in the scaling limit, thus creating a small (Lilliputian) world from the point  of view of the (Gulliver) lattice people.

If we do not renormalize the external lengths $L$ and $\beta$, it follows from 
\rf{newrho} that $\brho$ diverges in the scaling limit $\Lambda \to \infty$. This is 
a reflection of the fact that by integrating out the quantum fluctuations of $X^\mu_q$
 in the decomposition $X^\mu=X^\mu_{cl} + X^\mu_q$ the typical 
quantum world sheet will have an infinite area. However, the renormalization
\rf{LR} brings this back to a finite value in the limit $\Lambda \to \infty$ since 
we then obtain the metric
\be
\brho_R= \frac{L_R}{\omega_L\omega_\beta}
\frac{\left(\beta^2_R-\frac{\pi(d-2)}{6K_R}\right)}{\sqrt{\beta^2_R-\frac{\pi(d-2)}{3K_R}}}.
\label{newrhoR} 
\ee
Similarly, the renormalized mean-field ground state energy becomes finite 
\be
E_R = K_R  \sqrt{\beta^2_R-\frac{\pi(d-2)}{3K_R}},
\label{SfinR}
\ee
reproducing the well-known Alvarez-Arvis formula.
 
In the next Sections we will study the stability  of the mean-field vacuum \rf{mmff},
both when the ``lattice" renormalization and ``string theory" renormalization
are used.

\section{2D determinants and the Seeley expansion\label{ss:Seeley}}

Two-dimensional determinants diverge and have to be regularized. 
A standard regularization via the proper time is defined by
\bea
&&\log \det (- {\cal O})\Big|_{\rm reg}=
\tr\log (-{\cal O}) \Big|_{\rm reg}
= - \int_{a^2}^\infty \frac{\d\tau}\tau \tr  \e^{\tau {\cal O}}, \non
&&~~~~a^2 \equiv \frac1{4\pi \Lambda^2 }
\label{5}
\eea
with ${\cal O}$ given in \eq{aux1} or \eq{gh}.

The standard computation of the (proper-time regularized) determinants of 2D
operators
is based on the formula
\be
-\frac 12 \rho_{ab} (\omega) \frac {\delta}{\delta \rho_{ab}(\om)}
\tr\log (-{\cal O}) \Big|_{\rm reg}
=\LA \om |\e^{a^2{\cal O}} | \om\RA,
\label{rrrrrp}
\ee
where one substitutes the expansion  in $a^2$ of the matrix element of
the heat kernel operator
on the right-hand side, known as the Seeley expansion.
To two leading orders it is well-known \cite{deWitt,Gil75} for the bulk part.
The boundary terms are also known \cite{DOP82,Alv83} for 
our case of the Dirichlet (or Robin) boundary conditions, but we shall not need them
for $L\gg \beta$ so below we only write  the bulk terms.
We then have \cite{deWitt,Gil75,DOP82,Alv83} 
\bea
\LA \omega \Big| \e^{a^2 \rho^{-1} \partial _a \lambda^{ab}  \partial_b}
\Big| \omega \RA =
\frac1{4\pi a^2} \frac{\rho}{\lambda}+\frac{1}{4\pi}\left[-\frac 16 \partial_a^2 \ln \rho 
\right. \non \left.-
\frac 13 \partial_a^2 \ln \lambda - \frac 14 ( \partial_a \ln \lambda)^2
\right] +{\cal O}(a^2)
\label{Seeley}
\eea
for diagonal $\lambda^{ab}=\lambda \delta^{ab}$ and $\rho_{ab}=\rho\delta_{ab}$,
while the general case can be obtained by making use of
diffeomorphism invariance.

Given Eqs.~\rf{Seeley} and \rf{rrrrrp}, we can restore the effective action to quadratic order
in fluctuations,
except for the term $(\delta \lambda)^2$ whose variation with respect to $\rho$ vanishes.
We can directly compute this term (as well as the terms $(\delta \rho)^2$ and $\delta \rho\delta \lambda$) by the standard technique of calculating the Coleman-Weinberg potential 
to quadratic order. For this purpose we
expand the regularized determinant to the
second  order in $\delta \lambda$ (and $\delta\rho$) and obtain%
\bea
\lefteqn{A_{\lambda\lambda}(p)=\frac{1}{2\blambda^2} \int_{\blambda a^2/\brho}^{\infty} \d \tau
\int_0^\tau \d \sigma\,  \int \frac {d^2 k}{(2\pi)^2} }  \non
&&~~~\times k_a(k+p)_a\e^{-\sigma k^2 }(k+p)_bk_be^{-(\tau-\sigma)(k+p)^2}
\label{19}
\eea
for  the appropriate coefficient
of the quadratic form coming from the determinant.
Equation~\rf{19} is applicable in our case of one compactified dimension for $\beta \gg \sqrt{1/K_0}$. Otherwise,an additional (L\"uscher) term appears from the difference between
the integral and the discrete sum over $k_2$. It is explicitly written in \eq{SpS}
for constant $\lambda$ and $\rho$.

Integrating over $\sigma$ and $\tau$, we obtain
\bea
\lefteqn{A_{\lambda\lambda}(p)=\frac{1}{2\blambda^2}\int\frac {d^2 k}{(2\pi)^2}
[ k_a(k+p)_a]^2 } \non && \times
\left[ \frac {\e^{-a^2\blambda(k+p)^2/\brho}}{(k+p)^2}-\frac{\e^{-a^2\blambda k^2/\brho}}{k^2}\right]\frac1{(k+p)^2-k^2}.~~
\eea
Expanding in $a^2$, we then find 
\be
A_{\lambda\lambda}= -\frac\brho{4\pi a^2\blambda}-\frac {p^2}{16\pi}
\log \left(\frac{c \blambda p^2 a^2}\brho\right),
\label{66}
\ee
where $c$ is a (non-universal) constant. 

Analogously, for the ghost  determinant we have from the Seeley
expansion the standard result
\bea
\lefteqn{\tr \log 
\left\{\left[ -\frac 1{\rho} \partial^2 +
\frac 1{2\rho} (\partial^2 \log \rho) \right]\delta^a_b \right\}} \non && = 
-\Lambda^2 \int \d^2 \omega\, \rho
-\frac{13}{48 \pi} \int \d^2 \omega\,(\partial_a \log \rho)^2,
\label{gg}
\eea
where we write only the bulk term, so it does not depend on the boundary conditions.

Combining all together, we obtain
the effective action to quadratic order in fluctuations
\begin{widetext}
\bea
\delta  S_2&=&-\left(K_0-\frac{d\Lambda^2 }{2\blambda^2}\right) \brho \blambda
\!\!\int \d^2\omega \, 
\frac{\delta\rho}{\brho} \frac{\delta \lambda}{\blambda}
-\frac{d\Lambda^2\brho }{2\blambda} \!\!\int \d^2\omega \, 
\Big(\frac{\delta \lambda}{\blambda}\Big)^2
+\frac{(26-d)}{96\pi } \!\!\int \d^2 \omega \, \Big(\frac{\partial_a \delta \rho}{\brho}\Big)^2 \non &&
\!\!\!-\frac{d}{24\pi} \!\!\int \d^2 \omega \, \Big(\frac{\partial_a \delta \rho}{\brho}\Big) 
\Big(\frac{\partial_a \delta \lambda}{\blambda}\Big)
+\frac{ d }{32 \pi } 
\!\!\int \frac{\d^2 p }{(2\pi)^2}\,\Big(\frac{\delta \lambda(p)}{\blambda}\Big)
\Big(\frac{\delta\lambda(-p)}{\blambda}\Big)\;
p^2\log\Big(\frac{\Lambda^2 \brho}{c p^2 \blambda}\Big).
\label{S2}
\eea
\end{widetext}
Notice the last term on the right-hand side is normal  (and therefore regularization dependent) rather than anomalous as the third and fourth terms are.

\section{Stability of the effective action to quadratic order}

In the previous Section we have performed the computation
assuming that $\lambda^{ab}=\lambda \delta^{ab}$.
In order to justify this assumption, let us consider the divergent part
of the effective action for nondiagonal $\lambda^{ab}$
\bea
S_{\rm div}&=&\int\d^2 \omega
\left[ \frac{K_0}2 \lambda^{ab} \partial_a X_{\rm cl}\cdot \partial_b X_{\rm cl}+
K_0  \rho\left(1 -\frac12  \lambda^{aa} \right) \right.  \non&& \left .-
\frac {d \Lambda^2}2   \frac \rho{\sqrt{\det{\lambda}}}
+ \Lambda^2   \rho \right],  \quad  
\lambda^{aa}=\lambda^{11}+\lambda^{22}.
\label{cla}
\eea
The divergent part of \eq{SpS} above
 is the same as \eq{cla} for constant $\lambda^{ab}=\blambda \delta^{ab}$
and $\rho=\brho$.

Expanding to quadratic order 
\bea
&&\sqrt{ \det(\blambda \delta^{ab}+\delta \lambda^{ab})}=\lambda+\frac12 \delta \lambda^{aa}
-\delta \lambda_2 
+{\cal O}\left((\delta \lambda)^3\right), \non
&&\delta \lambda_2 =\frac1{8\blambda} (\delta \lambda_{11}-\delta \lambda_{22})^2+ 
\frac 1{2 \blambda}(\delta \lambda_{12})^2,
\label{3001}
\eea
we find from \rf{cla} for $\blambda=C$
\bea
S^{(2)}_{\rm div} &\!=\! &-\frac{d\Lambda^2 \brho}{2C} \int \d^2 \omega\,\delta \lambda_2
-\left(K_0-\frac{d\Lambda^2}{2C^2}\right)\!\int \d^2 \omega \, \delta \rho \frac{ \delta \lambda^{aa}}2
\non
&&-\frac{d\Lambda^2 \brho}{2C^3}\int \d^2 \omega \left( \frac{ \delta \lambda^{aa}}2\right)^2 .
\label{Sdi}
\eea

The first term on the right-hand side of \eq{Sdi} plays a very important role for dynamics
of quadratic fluctuations. Because the path integral over $\lambda^{ab}$ goes
 parallel to imaginary axis,
\ie  $\delta \lambda^{ab}$ is pure imaginary, 
 the first term is always {\em positive}. 
Moreover, 
its exponential plays the role of a (functional) delta-function as $\Lambda\to\infty$,
forcing $\delta \lambda^{ab}=\delta \lambda \,\delta^{ab}$.
The last two terms on the right-hand side of \eq{Sdi} then reproduce the first
two terms in \rf{S2}.

{}From \eq{S2} for the effective action 
to the second order  in fluctuations we find the following quadratic form:
\bea
\delta S_2&=&\int \frac{\d^2p}{(2\pi)^2} \left[
A_{\rho\rho} \frac{\delta \rho(p)\delta \rho(-p) }{\bar \rho^2} +
2A_{\rho\lambda} \frac{\delta \rho(p) \delta \lambda(-p)}{\bar  \rho \blambda} 
\right. \non && \left. \mbox{} \hspace*{1.5cm}
+A_{\lambda\lambda} \frac{\delta \lambda(p)\delta \lambda(-p) }
{\blambda^2} \right]
\label{191}
\eea
with
\begin{widetext}
\be
A_{ij} = \left[ \begin{array}{cc} 
\frac {(26-d) p^2}{96\pi}& -\frac12\left(K_0-\frac{d\Lambda^2 }{2C^2}\right)\brho C
-\frac {d p^2}{48\pi}\\-\frac12\left(K_0-\frac{d\Lambda^2 }{2C^2}\right)\brho C
-\frac {d p^2}{48\pi}
&-A
\end{array}
\right],
\label{MMMM}
\ee
\end{widetext}
where
\be
A=  \frac{d \Lambda^2\bar \rho} {2C} +\frac {d p^2}{32\pi}\log (cp^2/\Lambda^2 \brho) .
\label{AAAA}
\ee
For $p^2 \ll \Lambda^2 \bar \rho$, we can drop the second term on the right-hand side of
\eq{AAAA}, so $A$ becomes constant. 
For $p^2 \ga \Lambda^2 \brho$, $A$ depends on $p^2$ but
remains positive.

Since $\delta \lambda(\omega)$ is
pure imaginary, i.e.\ $\delta \lambda (- p) = - \delta\lambda^*(p)$, we find for the determinant associated with the matrix 
 in \eq{MMMM} 
\be
D= \left[\frac12\left(K_0-\frac{d\Lambda^2}{2C^2}\right) \brho C+
\frac {d p^2}{48\pi}\right]^2+ \frac {(26-d) p^2}{96\pi} A,
\label{D}
\ee
and the propagators corresponding to the action \rf{191} are given by 
\beq\label{xj3}
\LA \phi^*_i(p) \phi_j(p) \RA = \frac{A_{ij}}{D}, \qquad \phi_i =\left( \frac{\delta \rho}{\brho}
, \; 
\frac{\delta \lambda}{\blambda}\right). 
\eeq

For generic $K_0 > d \Lambda^2/2C^2$ the first term in \rf{D} dominates and we have a trivial stability of fluctuations for any  $d$: nothing propagates.
However,  we are really  interested  in the scaling regime \rf{sca}, 
where $K_R$ is finite as $\Lambda\to\infty$ and because of the scaling \rf{xj2} 
we now have two situations (1) lattice scaling where $\brho \sim \Lambda^2$ and (2)
string scaling where $\brho_R$ is finite according to \eq{newrhoR} for $\Lambda \to \infty$. 
In the latter case we can disregard the first term in
$D$ and the off-diagonal elements 
of the matrix  \rf{MMMM}, so that
\be
\frac{1}{\brho_R^2} \LA \delta \rho_R(p) \delta \rho_R(-p) \RA = \frac{48\pi} {(26-d) p^2}.
\label{32}
\ee
It is positive for $d<26$, but becomes negative for $d>26$ which may indicate a negative-norm
state. 
In the first case we obtain
\bea
\frac{1}{\brho^2} \LA \delta \rho(p) \delta \rho(-p) \RA &=& \frac{48\pi} {(26-d)}
\frac{1}{( p^2 + m^2)}, \non m^2 &\propto & \frac{K_R^2 \brho}{(26-d) d \Lambda^2},
\label{j32}
\eea
the mass being positive and finite as $\Lambda \to \infty$ for $d < 26$.

In both cases $\lambda$ stays  localized even in the
scaling limit, \ie $\lambda(\omega)=\blambda$. Thus only $\rho$ fluctuates.
This is similar to what is described in the
book~\cite{Pol87}.


\section{Discussion}

We have constructed the nonperturbative mean-field vacuum of the Nambu-Goto string
at finite $d$ disregarding fluctuations of $\rho_{ab}$ and $\lambda^{ab}$, which is an extension of the one \cite{AM15} at large $d$. We have demonstrated the
stability of this vacuum under fluctuations to quadratic order for $2<d<26$.

Because of the observed instability for $d>26$, a question arises as to how to
understand the expansion in fluctuations about the mean-field.
Originally, we expected
that it comes along with the expansion in $1/d$, like the $1/N$-expansion 
in the two-dimensional
$O(N)$ sigma-model. This would be indeed the case if $\delta \lambda$ 
was real, but in 
our case of imaginary $\delta \lambda$ the action is no longer stable for $d>26$.
We can still make sense of the expansion about the mean-field for $2<d<26$
as a semiclassical
WKB expansion about the nonperturbative ``classical'' vacuum, \ie that of an expansion
in the number of ``quantum'' loops. It is technically well-defined in the path-integral language
by assuming that the diagonal part $\delta \lambda $ is real at large $d$.

It is possible to compute such a ``quantum'' correction to the mean-field values of $C$ 
and of the energy of the string ground state. This should help to answer the
long-standing question of whether or not  the Alvarez-Arvis formula~\rf{SfinR}, which was
derived historically by the canonical quantization of the bosonic string with the Dirichlet boundary
conditions
and reproduced by our approach in
the mean-field approximation,
is exact not only at $d=26$ but also for $2<d<26$. 
The computation of such a correction will
involve only the propagator $\LA \delta \rho \delta \rho \RA$ 
 given in \eq{32}.
 It would be most interesting to
compute such a correction to the mean-field values.


{\bf Acknowledgments.} The authors acknowledge  support by  the ERC-Advance
grant 291092, ``Exploring the Quantum Universe'' (EQU).
Y.~M.\ thanks the Theoretical Particle Physics and Cosmology group 
at the Niels Bohr Institute for the hospitality.

\end{document}